\documentclass[twocolumn,letterpaper,secnumarabic,amssymb, nobibnotes, aps, prl]{revtex4-1}        
\usepackage{epsfig}
\usepackage{dcolumn}
\usepackage{epstopdf}
\usepackage{graphicx}
\begin{document}

\title{DFT quantization of the Gibbs Free Energy of a quantum body}  

 \author{S. Selenu}
\affiliation{}

\begin{abstract}
\noindent In this article it is introduced  a theoretical model made in order to perform calculations of the  quantum heat of a body that  could be acquired or delivered during a thermal transformation of its quantum states. Here the model is mainly targeted to the electronic  structure of matter\cite{Martin} at the nano and micro scale where  DFT models have been frequently developed of the total energy of quantum systems. Defining an Entropy functional $S[\rho]$ makes us able optimizing a free energy $G[\rho]$ of the quantum system at finite temperatures. Due to the generality of the model,  the latter  can be also applied to several first principles computational codes where ab initio modelling of quantum matter is asked.
\end{abstract}

\date{\today} 
\maketitle

 \section{Introduction}

\noindent  
This article will be  focused  on a study based on the derivation of  the quantum electrical heat absorbed or either released by a physical system made of atoms and electrons either at the nano or at the microscopic scale, extending to it the concept of thermal heat within a DFT model.  Also, the latter being  fundamental in physics, can be employed  for a full understanding of the thermal phase transformation of the electronic subsystem involved in the physics of crystals, where the thermal effects may directly affects their thermal stability while indirectly  enhancing their functionality, to it associated a lowering of the classical thermal gibbs free energy of electrons. Within this classical model shown here having a quantum counterpart we shall introduce in the next part of the article a comprehensive Density Functional Theory (DFT)\cite{Hohenberg,Sham,March,Parr} modelling taking into account the nature of  thermal quantum phenomena that has been avoided since the advent of DFT. A new DFT model based on  either a direct statistical interpretation of the electronic heat, as a thermodynamical degree of freedom, allows to  characterize the  energetics of the quantum body on  a base where a variational principle still  can be applied on  the searching of  the quantum electronic wave functions, either via computational simulations of quantum matter\cite{Martin}. In the following section it will be reported and developed a new DFT model of the quantum electronic heat while in the last section are reported  conclusions of the article.

\section{ DFT model of the electronic quantum heat}

Since 1960's the work on DFT modelling of quantum matter has been growing more until it becomes one of the most
diffused theoretical model employed for the understanding of the interactions of electrons in matter. Effort is mainly put  onto the development of this matter modelling in order showing for the first time  how to perform calculations of the quantum electronic heat, either absorbed or released by a quantum body during its thermal transformations at a finite temperature $T$. This work could then be considered as the  begining part of a more general theory can include an ab initio modelling of the first  law and second law of thermodynamics\cite{Fermi}, where heat and free energy variations of a  quantum body are taken into account. In fact it is actually considered a quantum system of electrons at a given temperature $T$ in thermal equilibrium with its surrounding. Electrons also are considered either interacting with themselves by a potential $V_{ee}$\cite{Sham},  an ionic external potential $V_{ie}$\cite{Sham},  having  also an exchange correlation potential\cite{Sham} $V_{xc}$ ruling their correlation, while an associated electronic charge density of the system is given by the following relation $\rho=\sum_n f_n \Psi^*_n\Psi_n$ times the electronic charge, i.e. a product of electronic elementary charge $e$ and of the quantum probability distribution $\rho$ of an event happening. The latter is calculated as the weighted average of the  products of wave functions of the quantum system at a state $n$, with  $f_n$ the number of electrons per state.  The  probability density can be directly used in order to perform routinely calculations of the potential energies usually encountered in quantum DFT,  considered then an eligible candidate for a further study of the quantum Gibbs free energy  and its related Entropy\cite{10,11,12,13,Shannon}. In order to base our statistics on a generalisation of the Shannon model of the Entropy, by optimizing at a variational level  a quantum Gibbs free energy functional,  a free energy operator is derived, and  related to a new set of differential eigenvalue equations. The latter, built as a functional of the electronic probability  density allows then deriving a set of  quantum differential equations by considering the very quantum mechanical interpretation  of the electronic density as  being the not normalized  probability of delocalized electrons being found in a position in space\cite{Ashcroft}. It is possible then recognizing to it being associated to a quantum Entropy whose expression is given by:

\begin{eqnarray}
\label{H2}
&&S=-k_B \int \rho ln \rho \\
&&S=-k_b\sum_n f_n \langle \Psi_n |ln \rho|\Psi_n\rangle
\end{eqnarray}

in agreement with the representation of the entropy of continuous distributions of probabilities as given by Shannon,
C. and Weaver, W.\cite{Shannon}. Here  a Lagrange multiplier $S=NlnN$ appears on variations of the energy functional, with $N=\int \rho$, being the latter the total  number of electrons  in the volume of the body. Latter result implies the writing of a new functional of the probability  density $\rho'={\rho}{N}$, that shall be  reported  in the next part of this section. By firstly  considering Gibbs free energy of the thermodynamical system begin given by the following classical  formula $G = H - TS $, where $Q = TS$  is the electronic heat at a given entropy $S$  for a fixed temperature$ T $ of the quantum system, an Enthalpy  functional $H[\rho]=E[\rho]+PV$ is calculated, being $E[\rho]$ the internal energy of the body. The  latter  energy functional, is  customarily  calculated  in quantum DFT and  kept the  volume $V$ of the body constant in order to neglect dependence of the energy of the system on the pressure $P$. The Enthalpy is then given by the sum of the internal energy plus a trivial constant $PV$ not varying with wave functions. The Gibbs  free energy functional  is then finally written making use of eq.(\ref{H2}) as it follows:

\begin{eqnarray}
\label{H02} 
G[N\rho']=H[N\rho']-NTS[\rho']+k_B T N ln N
\end{eqnarray}

made by the sum of the Helmholtz free energy\cite{Fermi,Enthalpy} and the electronic heat of the body. In what follows it is not varied only the Helmholtz free energy, where the product $PV$ is kept constant  in order calculating the Gibbs free energy variation, but also the Entropy functional. A  stationary differential equation is reached  and directly interpreted as a quantum mechanical eigenvalue equation whose eigen values $\epsilon_G$ are the eigen free energies of the quantum body. The functional derivatives of $G$, $H$ and $S$ are calculated by  referring to  eq.(\ref{H02}), obtaining  the following result:

\begin{eqnarray}
\label{H3}
&&\hat{G}\Psi_n=\epsilon_{n,G}\Psi_n \\\nonumber 
&&\hat{G}=[\frac{-\hbar^2 \nabla^2}{2m}+ V_{ee}+ V_{ie}+ V_{XC}+ k_BT ln \rho' + k_BT]
\end{eqnarray}

An Hamiltonian eigenvalue set of equations can also be reached by the shifting  of energies $\epsilon_n=\epsilon_{n,G}+k_BT$, due to the increase of the latter of a thermal energy amount equal to $ k_BT$, related to  the thermal motion of electrons,  allowing to reach the following stationary equation:

\begin{eqnarray}
\label{H03}
&&\hat{H}\Psi_n=\epsilon_n \Psi_n \\\nonumber 
&&\hat{H}=[\frac{-\hbar^2 \nabla^2}{2m}+ V_{ee}+ V_{ie}+ V_{XC}+ TV_S ]
\end{eqnarray}

It appears a new thermodynamical potential on right hand side of equation (\ref{H03}) directly dependent on the normalized probability density,  $V_S=k_B ln \rho'$, whose expectation values with respect the $nth$ state of the electronic wave $\Psi_n$ is the Entropy per electron $S'$  of the system so as to call it  the entropic potential. It becomes evident  we shall add the latter to the new DFT operator, reported in eq.(\ref{H3}), in order calculating the eigenstates of quantum matter in its thermal states.  Due to the definition of the quantum electronic heat, being $Q = TS$, it is straightforward showing by eq.(\ref{H2}) the expression of the electronic heat to be  written as follows:

\begin{eqnarray}
\label{H003}
Q=-Nk_BT\int \rho' ln \rho'
\end{eqnarray}
with respect to the electronic probability density $\rho'$, making the free energy a minimum as it is the natural tendency of a physical system due to the second law of thermodynamics\cite{Enthalpy,Fermi}. This has been the first attempt to study the quantum thermodynamics of matter in its steady states in a Hamiltonian formalism, brought us introducing the quantum free energy operator,  then  calculate eigenstates along any thermal  changings of the quantum state of matter. Having reached for the first time  a quantization of the Gibbs free energy   the article will be concluded in the next section.

\section{Conclusions}
This article is concluded having shown how to calculate the quantum electronic heat of a body, at a DFT variational level, also defining the electronic Entropy functional  S[$\rho$], it interpreted as the Entropy of the system. Latter result  makes us able optimizing the quantum free energy G[$\rho$] by variations of the eigenstates of the quantum system at finite temperature $T$. Due to the generality of our model we have reached a  set of differential equations  being expressed in (\ref{H3}), showing that it can be applied to several first principle codes of an ab initio modelling of matter, either at the nano or  microscopic scale. A future computational modelling of the electronic heat delivered or either absorbed by a quantum system during its  thermal phase transformations can then be targeted.

\end{document}